# Machine learning to optimize precision in the analysis of randomized trials: A journey in pre-specified, yet data-adaptive learning

*Authors:* Laura B. Balzer, Mark J. van der Laan, Maya L. Petersen

*Affiliation:* Division of Biostatistics, School of Public Health, University of California Berkeley, Berkeley, CA

*Corresponding author:* Laura B. Balzer, Division of Biostatistics, School of Public Health, University of California Berkeley, 2121 Berkeley Way West, Berkeley, CA 94720, USA; laura.balzer@berkeley.edu

*Funding*: This work was supported, in part, by the National Institutes of Health (NIH) under award number U01AI150510 and a philanthropic gift from the Novo Nordisk corporation to the University of California, Berkeley for the Joint Initiative for Causal Inference (JICI). The content is solely the responsibility of the authors and does not necessarily represent the official views of the NIH or Novo Nordisk.

*Trial registrations:* NCT01864603, NCT04810650, NCT05549726 at ClinicalTrials.gov.

*Abstract*:

Covariate adjustment is an approach to improve the precision of trial analyses by adjusting for baseline variables that are prognostic of the primary endpoint. Motivated by the SEARCH Universal HIV Test-and-Treat Trial (2013-2017), we tell our story of developing, evaluating, and implementing a machine learning-based approach for covariate adjustment. We provide the rationale for as well as the practical concerns with such an approach for estimating marginal effects. Using schematics, we illustrate our procedure: targeted machine learning estimation (TMLE) with Adaptive Pre-specification. Briefly, sample-splitting is used to data-adaptively select the combination of estimators of the outcome regression (i.e., the conditional expectation of the outcome given the trial arm and covariates) and known propensity score (i.e., the conditional probability of being randomized to the intervention given the covariates) that minimizes the cross-validated variance estimate and, thereby, maximizes empirical efficiency. We discuss our approach for evaluating finite sample performance with parametric and plasmode simulations, pre-specifying the Statistical Analysis Plan, and unblinding in real-time on video conference with our colleagues from around the world. We present the results from applying our approach in the primary, pre-specified analysis of 8 recently published trials (2022-2024). We conclude with practical recommendations and an invitation to implement our approach in the primary analysis of your next trial.

Imagine you are tasked with pre-specifying and implementing the primary analysis of a population-level randomized trial. Concretely, you are responsible for evaluating effectiveness in the SEARCH Universal HIV Test-and-Treat Trial, which was conducted from 2013 to 2017 in Kenya and Uganda (NCT01864603).(1) Based on standard sample size calculations for cluster-randomized controlled trials,(2) you anticipate that 32 communities (each with ~10,000 persons) will provide sufficient power for the primary endpoint of HIV incidence, a rare binary outcome. Given the randomization and follow-up procedures, you know the unadjusted effect estimator will be unbiased. However, you have also heard promises about the potential gains in precision and power with covariate adjustment: an analytic approach to adjust for baseline variables that are prognostic of the outcome.(3,4) You may have seen promising results from simulation studies for a wide range of causal estimands and outcomes.(5–12) You may be particularly intrigued by targeted machine learning estimation (TMLE), which is a model-robust, locally efficient approach and illustrated in **Figure 1** (**eAppendices A-B**).(13,14)

However, you may also be troubled by practical concerns – at least we were… How do we *a priori* specify the optimal adjustment approach? What if we bet incorrectly and implement an adjusted analysis that actually harms precision? What if the adjusted analysis gives wildly different results from the unadjusted one? How will the procedure perform in small sample settings – we only have 32 independent units? How will the procedure perform with a rare binary outcome? Is there a risk of under-estimating the variance and inflated Type-I error rates? Will stakeholders understand and trust the results? Overall, we needed a method that was data-adaptive, yet fully pre-specified *and* maintained excellent performance.

Due to treatment randomization, trials are highly robust for conventional intention-to-treat analyses with minimal outcome missingness. This setting allows us to focus our analytic efforts on minimizing variance. Therefore, building on the theory of Empirical Efficiency

Maximization,(15,16) we proposed a practical and automated procedure using machine learning to select the optimal adjustment approach in randomized trials. Given the procedure is both pre-specified and adaptive, we called it "Adaptive Pre-specification".(7)

For implementation of this procedure with TMLE, we must pre-specify:

1. Candidate estimators of the "outcome regression", which is the expected outcome given the trial arm and covariate(s).
2. Candidate estimators of the "propensity score", which is the conditional probability of being randomized to the intervention given the covariates(s). Although this probability is known in trials, estimating it can improve precision.(17)
3. Cross-validation procedure, including the process for splitting the data, evaluating the candidates, and obtaining final point and variance estimates.
4. Loss function to quantify performance of the candidates. We use the estimated influence curve (function)-squared, which provides a variance estimate for each candidate (pg 572-577 in (13)).

Altogether, the "best" estimator is the candidate minimizing the cross-validated variance estimate and, thereby, maximizing empirical efficiency. The unadjusted estimator *must* be included as a candidate for the outcome regression and as candidate for the propensity score so that it can be selected if none of the candidates using covariate adjustment improves precision.

As illustrated in **Figure 2**, we use sample-splitting to first select the outcome regression estimator that minimizes the cross-validated variance estimate. Then using the selected outcome regression estimator, we select the propensity score estimator that *further* minimizes the cross-validated variance estimate. Together, the selected outcome regression estimator and propensity score estimator form the optimal TMLE, which is guaranteed to be at least as precise as the unadjusted effect estimator.

Then the question was how does TMLE with Adaptive Pre-specification perform in practice and, most importantly, in settings mirroring our trial? To assess performance across a variety of scenarios, we conducted parametric simulations, which demonstrated that our procedure offered meaningful gains in power while maintaining nominal confidence coverage.(7) To assess expected performance for our trial, we conducted plasmode simulations, which sample partially from the empirical distribution.(9,18–20) Specifically, we sampled baseline covariates from the SEARCH data, randomized the trial arms according to the study design, and simulated the outcome using mathematical models of the HIV epidemic.(21) These simulations demonstrated that our adaptive procedure improved power while preserving confidence interval coverage when there was an effect and protected Type-I error control under the null.(6)

Finally, we were ready to pre-specify the Statistical Analysis Plan (SAP) and analytic code to implement TMLE with Adaptive Pre-specification in the primary analysis of SEARCH.(1,21,22) The SAP and code were locked prior to outcome measurement and while the primary trial statistician (LBB) remained blinded. As a final check, we conducted treatment-blind plasmode simulations to confirm Type-I error control. We loaded in the actual SEARCH data, permuted the arm labels between clusters (the randomized unit), implemented our primary analytic approach, and repeated many times – verifying the proportion of times that the true null hypothesis was rejected was ≤5%.

Then on April 6, 2018, we unblinded. Live on video conference with our colleagues from around the world, the primary statistician (LBB) shared her screen, loaded the actual data into *RStudio*, and executed the code. Together and in real-time, we learned not only the effectiveness of the intervention, but also which variables were selected for covariate adjustment (if any). The primary analyses, using this machine learning approach, were published in the *New England Journal of Medicine*.(1)

The efficiency gains achieved in SEARCH have previously been published.(11) For HIV incidence (primary endpoint), the primary analysis – TMLE with Adaptive Pre-specification – was nearly 5-times more precise than the unadjusted effect estimator (the simple contrast in arm-specific average outcomes). Specifically, the estimated variance of TMLE was nearly 1/5th that of the unadjusted approach. For the secondary endpoints of tuberculosis incidence and hypertension control, TMLE was 2.6 and 1.8 times more precise, respectively. For HIV viral suppression, another secondary endpoint, we saw no difference in precision. This range of precision gains – from substantial to none, but with no instances of harm – is a key characteristic of our approach. Our procedure defaults to the unadjusted estimator when none of the pre-specified candidates improves empirical efficiency. We are protected from forced adjustment to the detriment of precision.

Given our motivation to maximize power in a trial with only 32 randomized units and a rare outcome, we originally limited the candidate estimators to "working" generalized linear models (GLMs) adjusting for at most 1 covariate.(6) Recently, we extended the procedure for larger trials and to include "well-behaved" machine learning algorithms as candidates: stepwise regression, penalized regression, and multivariate adaptive regression splines with and without screening.(12) These approaches respond flexibly to the data, but are less likely to overfit (i.e., they satisfy the "Donsker" conditions).(13,23) Across a variety of settings, our simulations demonstrated meaningful precision improvements – translating into 20-43% reductions in sample size for the same statistical power.(12) (These sample size savings were calculated as 1 minus the mean squared error [MSE] of a covariate-adjusted estimator divided by the MSE of the unadjusted estimator.(9)) Importantly, precision gains were achieved for both binary and continuous endpoints and without sacrificing 95% confidence interval coverage. These simulations also highlighted the precision gains with data-adaptive adjustment, versus forced adjustment for the variables used in stratified randomization. As a real-data demonstration, we re-analyzed ACTG Study 175 and found the selected TMLE varied by pre-specified subgroups – highlighting the importance of avoiding a "one-size-fits-all" approach.(12)

Since its publication in 2016, we have applied TMLE with Adaptive Pre-specification in the primary analyses of over a dozen trials.(1,24–35) In **Table 1**, we present the precision gains and optimal adjustment approach for 8 recently published trials (NCT04810650; NCT05549726). As before, there was a range of efficiency gains, but crucially, we never lost precision. Notably, within and across trials, the selected TMLE varied – highlighting the flexibility of our pre-specified, yet data-adaptive approach. In all cases and as expected, the point estimates from the TMLE and the unadjusted estimator were very similar (**eAppendix C**).

Reflecting the motivating studies (**Table 1**), we have focused on trials with intent-to-treat estimands and ignorable missingness. Other trials may be subject to differential missingness or censoring, which can cause meaningful bias. For individually randomized trials, various TMLEs have been developed and applied to address these common complications.(8–10,36,37) For cluster randomized trials, we developed and applied Two-Stage TMLE to first account for missingness and other potentially biasing factors at the individual-level and then to evaluate effectiveness with maximum precision by applying a cluster-level TMLE with Adaptive Pre-specification.(1,11,24,25,38–41)

While our real-data analyses demonstrate the value of our approach, skepticism remains. We are often asked about the rationale for using an adaptive approach when the team "knows" the most prognostic covariate. As illustrated by our prior analyses, it is difficult to bet *a priori* on the best adjustment covariate(s) and the form of the working regressions. Furthermore, the optimal approach may vary by subgroup or endpoint. Therefore, we recommend letting the data decide through our fully pre-specified, yet adaptive procedure. We offer similar advice to teams who have built a prognostic score based on historical data and plan to adjust for that score in their analysis (a.k.a., "PROCOVA").(42,43) While this score may effectively summarize the covariate set into a single measure, it may not be *the* optimal adjustment covariate. Therefore, we recommend including it as another candidate in our procedure.

We emphasize that our algorithm requires detailed pre-specification of the primary estimand (implying a specific loss function – the estimated influence curve-squared), the candidate approaches for estimating the outcome regression and propensity score, and the cross-validation procedure (**eAppendix C**). Conducting simulations, imitating the key features of the trial, can help inform these choices and confirm implementation of the computing code.(20,21) While recognizing each trial offers unique opportunities and challenges, we offer the following recommendations. First, if you have concerns about data sparsity (e.g., due to rare outcomes or few independent units), limit the candidate adjustment algorithms to a handful of working GLMs adjusting for at most 1 candidate. Second, if you have concerns about perceived or actual overfitting, use the cross-validated variance estimate, which is a natural by-product of the algorithm (**Figure 2**). In our experience, the cross-validated variance estimate has been overly conservative, but this may not always be the case, especially when using more aggressive learners. Finally, we reiterate that the unadjusted estimator *must* always be included as a candidate so that it can be selected if none of the candidate approaches using covariate adjustment improves precision. (See **eAppendix D** for implications for power calculations.)

We also emphasize the importance of reproducibility and transparency. First, in line with best practices, the analytic code must be pre-specified. When using data-adaptive methods, such as our approach, we must include additional safeguards, including setting-the-seed and documenting the software version. Second, we have thoroughly appreciated the process of unblinding and running the primary analysis in real-time on video conference. However, we recognize this might not be possible for all studies. At minimum, we recommend using *RMarkdown* (or a similar markup language) to generate a user-friendly document with time-stamped results for sharing immediately after unblinding. Finally, we recommend using TMLE with Adaptive Pre-specification in the primary analysis and including the details of the selected approach and comparisons to the unadjusted analysis in supplementary materials.

In conclusion, TMLE with Adaptive Pre-specification is fully pre-specified, model-robust, and automated procedure to data-adaptively select the adjustment approach that maximizes empirical efficiency in randomized trials. It is theoretically supported, guaranteed not to harm precision, and widely applicable. Prior use meaningfully improved precision in several high-profile trials, which were published in top-tier journals. Covariate adjustment is endorsed by both the U.S. Food and Drug Administration as well as the European Medicines Agency.(3,44,45) Computing code to implement TMLE with Adaptive Pre-specification is publicly available.(46) More precise analyses translate to reduced uncertainty, improved power, and fewer Type-II errors. So perhaps you will consider applying our approach in your next analysis?

**Acknowledgements:** We gratefully thank the leaders of and our collaborators on the Sustainable East Africa Research in Community Health (SEARCH) Consortium (https://www.searchendaids.com/): Diane V. Havlir, Moses R. Kamya, James Ayieko, Jane Kabami, Elijah Kakande, Gabriel Chamie, Matthew Hickey, and many others who contributed to the trials motivating this work. The trials reported in Table 1 were supported by the NIH (U01AI150510), and ViiV Healthcare provided cabotegravir long-acting in one trial. The content is solely the responsibility of the authors and does not necessarily represent the official views of the NIH or ViiV. On behalf of the SEARCH team, we also thank the Ministry of Health of Uganda and of Kenya; our research teams and administrative teams in San Francisco, Berkeley, Uganda, and Kenya; our advisory boards, and especially all communities and participants involved.

**Declaration of Conflicting Interests:** The Authors declare that there is no conflict of interest.

**Figure 1:** Schematic of steps to obtain a point and variance estimate with TMLE for the marginal effect defined as contrast of the expected counterfactual outcomes. The targeting procedure and the form of the influence curve will depend on the estimand. See eAppendix A for further details.

**A)** Regress the outcome $Y$ on the trial arm $A$ and adjustment covariates $W$. Then use the regression fit to obtain "initial predictions" of the expected outcome for all observations under the intervention $A = 1$ and under the control $A = 0$.

| Id | Covariates | | | Arm | Outcome | Initial predictions | |
|---|---|---|---|---|---|---|---|
| 1 | $W1_1$ | $W2_1$ | $W3_1$ | $A_1$ | $Y_1$ | $\widehat{\mathbb{E}}(Y\|A=1,W_1)$ | $\widehat{\mathbb{E}}(Y\|A=0,W_N)$ |
| | | | | | | | |
| N | $W1_N$ | $W2_N$ | $W3_N$ | $A_N$ | $Y_N$ | $\widehat{\mathbb{E}}(Y\|A=1,W_N)$ | $\widehat{\mathbb{E}}(Y\|A=0,W_N)$ |

**B)** Regress the trial arm $A$ on the adjustment covariates $W$. Then use the regression fit to obtain estimates of the propensity score ("Pscore") for all observations.

| Id | Covariates | | | Arm | Outcome | Initial predictions | | Pscore ests. |
|---|---|---|---|---|---|---|---|---|
| 1 | $W1_1$ | $W2_1$ | $W3_1$ | $A_1$ | $Y_1$ | $\widehat{\mathbb{E}}(Y\|A=1,W_1)$ | $\widehat{\mathbb{E}}(Y\|A=0,W_N)$ | $\widehat{\mathbb{P}}(A=1\|W_1)$ |
| | | | | | | | | |
| N | $W1_N$ | $W2_N$ | $W3_N$ | $A_N$ | $Y_N$ | $\widehat{\mathbb{E}}(Y\|A=1,W_N)$ | $\widehat{\mathbb{E}}(Y\|A=0,W_N)$ | $\widehat{\mathbb{P}}(A=1\|W_N)$ |

**C)** Regress the outcome $Y$ on the "clever covariate" (a function of the propensity score estimates) with the initial outcome predictions as offset. Using the updated regression fit, obtain "targeted predictions" of the expected outcome for all observations under the intervention $A = 1$ and under the control $A = 0$.

| Id | Covariates | | | Arm | Outcome | Initial predictions | | Pscore ests. | Targeted predictions | |
|---|---|---|---|---|---|---|---|---|---|---|
| 1 | $W1_1$ | $W2_1$ | $W3_1$ | $A_1$ | $Y_1$ | $\widehat{\mathbb{E}}(Y\|A=1,W_1)$ | $\widehat{\mathbb{E}}(Y\|A=0,W_N)$ | $\widehat{\mathbb{P}}(A=1\|W_1)$ | $\widehat{\mathbb{E}}^*(Y\|A=1,W_1)$ | $\widehat{\mathbb{E}}^*(Y\|A=0,W_N)$ |
| | | | | | | | | | | |
| N | $W1_N$ | $W2_N$ | $W3_N$ | $A_N$ | $Y_N$ | $\widehat{\mathbb{E}}(Y\|A=1,W_N)$ | $\widehat{\mathbb{E}}(Y\|A=0,W_N)$ | $\widehat{\mathbb{P}}(A=1\|W_N)$ | $\widehat{\mathbb{E}}^*(Y\|A=1,W_N)$ | $\widehat{\mathbb{E}}^*(Y\|A=0,W_N)$ |

**D)** Obtain a point estimate of the marginal effect by averaging the targeted predictions in each arm and contrasting.

$$\widehat{\Psi}^* = \frac{1}{N}\sum_{i=1}^{N} \widehat{\mathbb{E}}^*(Y|A=1,W_i) \quad \text{versus} \quad \frac{1}{N}\sum_{i=1}^{N} \widehat{\mathbb{E}}^*(Y|A=0,W_i)$$

**E)** Obtain a variance estimate with the sample variance of the estimated influence curve $IC$, scaled by sample size $N$.

$$\widehat{\sigma}^2 = \frac{1}{N}\widehat{Var(IC)}$$

For example, if our estimand is the average treatment effect, the estimated $IC$ for observation i is $\widehat{IC}_i = \left(\frac{1(A_i=1)}{\widehat{\mathbb{P}}(A=1|W_i)} - \frac{1(A_i=0)}{\widehat{\mathbb{P}}(A=0|W_i)}\right)\left(Y_i - \widehat{\mathbb{E}}^*(Y|A_i,W_i)\right) + \widehat{\mathbb{E}}^*(Y|A=1,W_i) - \widehat{\mathbb{E}}^*(Y|A=0,W_i) - \widehat{\Psi}^*$.

**Figure 2**: Schematic of TMLE with Adaptive Pre-specification – using an example with 3 candidates for the outcome regression, 3 candidates for the propensity score, and 5-fold cross-validation. In Step J, we can alternatively use a cross-validated variance estimate, obtained as the sample variance of the cross-validated influence curve (IC)-estimates scaled by sample size. The procedure should only be implemented after fully pre-specifying the estimand (implying the loss function), the candidate estimators of the outcome regression, the candidate estimators of the propensity score, and the cross-validation procedure (including the approach for variance estimation).

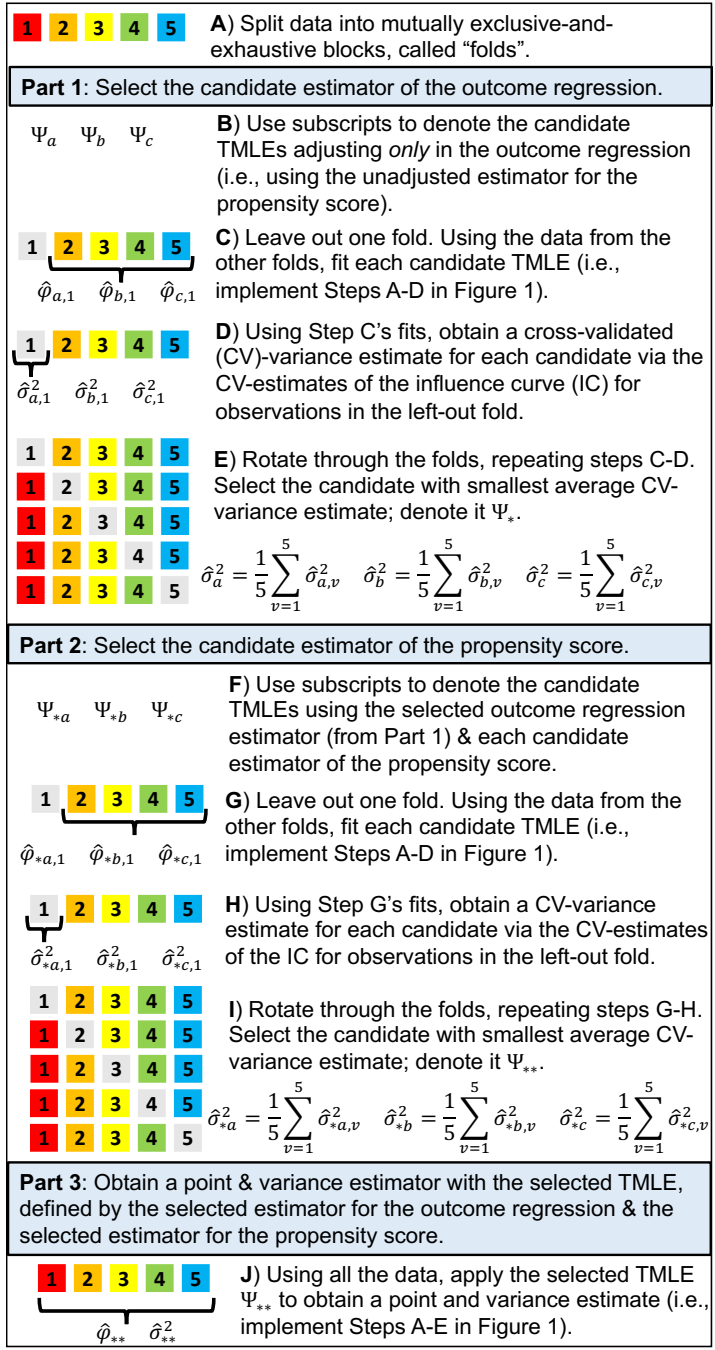

**Table 1:** Showcase of recently published trials where TMLE with Adaptive Pre-specification was used in the primary analysis (NCT04810650; NCT05549726). For the primary endpoint (first row) and a secondary endpoint (second row) of each trial, we show the precision gain (the estimated variance of the unadjusted approach divided by that of TMLE) and the selected approaches for estimating the outcome regression and the propensity score. Additional details are in eAppendix C.*

| Citation | Study population | Endpoint | Precision gain | Outcome regression | Propensity score |
|---|---|---|---|---|---|
| *Brief alcohol counselling (intervention) vs. the standard-of-care (control)* | | | | | |
| Puryear, et al. *JIAS*, 2023 | 401 persons (18+ years) with HIV & unhealthy alcohol use | Viral suppression | 1.11 | GLM adjusting for prior suppression | GLM adjusting for prior alcohol use |
| | | Alcohol use | 1.64 | Stepwise | MARS |
| *Dynamic choice strategy for HIV care (intervention) vs. the standard-of-care (control)* | | | | | |
| Ayieko, et al. *JAIDS*, 2024 | 201 highly mobile adults with HIV | Viral suppression | 1.11 | MARS with screening | Unadjusted |
| | | Treatment possession | 1.25 | GLM adjusting for prior suppression | GLM adjusting for prior mobility |
| *Financial incentive (intervention) vs. standard referral (control) following community-based screening* | | | | | |
| Hickey, et al. *PLoS One,* 2022. | 199 persons (25+ years) with uncontrolled hypertension | Linkage to clinic for care | 1.03 | GLM adjusting for site | Unadjusted |
| | | Hypertension control | 1.04 | GLM adjusting for baseline severity | Unadjusted |
| *Community health worker-facilitated telehealth (intervention) vs. clinic-based care (control)* | | | | | |
| Hickey, et al. *PLoS Med,* 2025 | 200 older adults (40+ years) with moderate-severe hypertension | Hypertension control | 1.12 | Main terms GLM | GLM adjusting for age |
| | | Systolic blood pressure | 1.18 | Main terms GLM | Unadjusted |
| *Dynamic choice HIV prevention strategy (intervention) vs. the standard-of-care (control)* | | | | | |
| Kakande, et al. *JIAS*, 2023 | 429 adults with HIV risk, recruited from the community | Prevention use | 1.39 | GLM adjusting for prior prevention use | Unadjusted |
| | | Use during risk periods | 1.12 | GLM adjusting for prior prevention use | Unadjusted |
| Kabami, et al. *JAIDS*, 2024 | 400 adult women with HIV risk, recruited from antenatal clinics | Prevention use | 1.14 | Stepwise | Unadjusted |
| | | Use during risk periods | 1.16 | Stepwise | GLM adjusting for country |
| Koss, et al., *AIDS*, 2024 | 403 adults with HIV risk, recruited from the outpatient department | Prevention use | 1.07 | GLM adjusting for prior prevention use | Unadjusted |
| | | Use during risk periods | 1.05 | GLM adjusting for prior prevention use | Unadjusted |
| *Dynamic choice HIV prevention strategy with injectable prevention (intervention) vs. the standard-of-care (control)* | | | | | |
| Kamya, et al. *Lancet HIV*, 2024 | 984 adults with HIV risk, recruited from clinics & the community | Prevention use | 1.23 | GLM adjusting for sex | GLM adjusting for prior alcohol use |
| | | Use during risk periods | 1.29 | GLM adjusting for prior alcohol use | GLM adjusting for age |

Abbreviations: Multivariate adaptive regression splines (MARS); generalized linear model (GLM)
*All trials took place in rural Kenya and Uganda. "Adult" refers to persons aged 15+ years.

27. Kabami J, Kabageni S, Koss CA, Okiring J, Nangendo J, Ruhamyankaka E, et al. A Peer-Mother Counseling Intervention Improves Early Infant HIV Testing in Rural Uganda. Pediatr Infect Dis J. 2025 July 22;

28. Kabami J, Balzer LB, Atukunda M, Arinaitwe E, Mutungi G, Twinamatsiko B, et al. A Multi-Component Integrated HIV and Hypertension Care Model Improves Hypertension Screening and Control in Rural Uganda: A Cluster Randomized Trial [Internet]. Rochester, NY: Preprints with The Lancet; 2024 [cited 2025 Aug 28]. Available from: https://papers.ssrn.com/abstract=5050332

29. Chamie G, Balzer L, Litunya J, Beesiga B, Marson K, others. A cluster randomized trial of multi-disease vs HIV-focused recruitment strategies to promote biomedical HIV prevention uptake among adults at alcohol-serving venues in rural Kenya and Uganda. In: International AIDS Society Conference. Rwanda; 2025.

30. Puryear SB, Mwangwa F, Opel F, Chamie G, Balzer LB, Kabami J, et al. Effect of a brief alcohol counselling intervention on HIV viral suppression and alcohol use among persons with HIV and unhealthy alcohol use in Uganda and Kenya: a randomized controlled trial. J Int AIDS Soc. 2023 Dec 6;26(12):e26187.

31. Ayieko J, Balzer LB, Inviolata C, Kakande E, Opel F, Wafula EM, et al. Randomized Trial of a "Dynamic Choice" Patient-Centered Care Intervention for Mobile Persons With HIV in East Africa. J Acquir Immune Defic Syndr 1999. 2024 Jan 1;95(1):74–81.

32. Kakande ER, Ayieko J, Sunday H, Biira E, Nyabuti M, Agengo G, et al. A community-based dynamic choice model for HIV prevention improves PrEP and PEP coverage in rural Uganda and Kenya: a cluster randomized trial. J Int AIDS Soc. 2023 Dec 6;26(12):e26195.

33. Kabami J, Koss CA, Sunday H, Biira E, Nyabuti M, Balzer, LB, et al. Randomized trial of dynamic choice HIV prevention at antenatal and postnatal care clinics in rural Uganda and Kenya. JAIDS. 2024;95(5):447–55.

34. Koss CA, Ayieko J, Kabami J, Balzer LB, Kakande E, Sunday H, et al. Dynamic choice HIV prevention intervention at outpatient departments in rural Kenya and Uganda: a randomized trial. AIDS. 2024;38(3):339–49.

35. Kamya MR, Balzer LB, Ayieko J, Kabami J, Kakande E, Chamie G, et al. Dynamic choice HIV prevention with cabotegravir long-acting injectable in rural Uganda and Kenya: a randomised trial extension. Lancet HIV. 2024 Nov 1;11(11):e736–45.

36. Moore KL, van der Laan MJ. Increasing Power in Randomized Trials with Right Censored Outcomes Through Covariate Adjustment. J Biopharm Stat. 2009;19(6):1099–131.

37. Maringe C, Smith MJ, Phillips RV, Luque-Fernandez MA. Application of targeted maximum likelihood estimation in public health and epidemiological studies: a systematic review. Ann Epidemiol. 2023 Oct 1;86:34-48.e28.

38. Hickey MD, Ayieko J, Owaraganise A, Sim N, Balzer LB, Kabami J, et al. Effect of a patient-centered hypertension delivery strategy on all-cause mortality: Secondary analysis
18

**SUPPLEMENTARY MATERIALS**

**Machine learning to optimize precision in the analysis of randomized trials: A journey in pre-specified, yet data-adaptive learning**

Laura B. Balzer, Mark J. van der Laan, Maya L. Petersen

**eAppendix A: Additional Details on TMLE in Randomized Controlled Trials (RCTs)**

In Figure 1 of the main text, we provide a schematic outlining how to obtain a point and variance estimate for the marginal effect (i.e., contrast in expected counterfactual outcomes) with targeted machine learning (TMLE) in randomized trials.(1,2) Here, we provide a step-by-step tutorial. For each step, we provide sample *R* code. Full computing code to implement TMLE with Adaptive Pre-specification is provided in https://github.com/LauraBalzer/AdaptivePrespec. Without loss of generality, we focus on a binary or bounded continuous outcome; as detailed in Gruber and van der Laan, we scale the outcome to be [0,1], implement the below procedure, and then scale-back.(3)

0. <u>Create a dataset with the observed covariates, trial arm, and outcome.</u> Create two analogous datasets where all observations receive the intervention and all observations receive the control.

```
df <- data.frame(W1=W1, W2, A, Y)
# copying
df_1 <- df_0 <- df
# setting A=1 for all
df_1$A <-1
# setting A=0 for all
df_0$A <- 0
```

1. <u>Obtain initial predictions of the expected outcomes under the intervention $\hat{\mathbb{E}}(Y|A=1,W)$ and under the control $\hat{\mathbb{E}}(Y|A=0,W)$.</u> Regress the outcome *Y* on the arm *A* and adjustment covariates *W* according to a "working" generalized linear model (GLM) with the logit link.(2) Throughout, "working" refers to the fact that this regression need not be and often is not correctly specified. We are only adjusting for prognostic covariate(s) to improve precision – not eliminate bias. Then using the regression fit, obtain the expected outcome for all observations under the observed trial arm, the intervention *A=1*, and the control *A=0*.

```
# initial fit of the outcome regression according to working GLM
# here we demonstrate adjustment for W1 in a pre-specified main terms model
initial_fit <- glm(Y~A + W1, family="binomial")
```

# initial predictions under the observed arm
# throughout, `type="response"` ensures that we get predictions on the relevant scale
initial_prediction_A <- predict(initial_fit, newdata=df, type="response")
# initial predictions under the treatment; here df_1 is the data frame with A=1 for all
initial_prediction_1 <- predict(initial_fit, newdata=df_1, type="response")
# initial predictions under the control; here df_0 is the data frame with A=0 for all
initial_prediction_0 <- predict(initial_fit, newdata=df_0, type="response")

2. <u>Estimate of the propensity score $\mathbb{P}(A = 1|W)$ for all observations</u>. In a trial, the propensity score is known, but estimating it can further improve precision.(1,4). Then using the regression fit, predict the propensity score for all observations.

# here, we demonstrate adjustment for W2 in a pre-specified main terms model
pscore_fit <- glm(A ~ W2, family="binomial")
# predict the propensity for all observations
# again, we use `type="response"` to ensure that predicted probabilities are returned
pscore <- predict(pscore_fit, type="response")

3. <u>Obtain targeted predictions of the expected outcome for all observations under the intervention $\widehat{\mathbb{E}}^*(Y|A = 1, W)$ and under the control $\widehat{\mathbb{E}}^*(Y|A = 0, W)$.</u> The exact form of the targeting step will depend on the estimand of interest. The implementation for effect on the absolute scale (i.e., average treatment effect or risk difference) has been covered extensively (e.g., (5,6)). Here, we demonstrate the updating step through a two-dimensional "clever covariate" to target effects on the absolute or relative scale.(1) First, we calculate the clever covariate for each arm as an indicator of being randomized to that arm, divided by the propensity of being randomized to that arm: $\frac{1(A=a)}{\mathbb{P}(A=a|W)}$ for A={1,0}. Then we regress the observed outcomes on the clever covariates with the (logit of the) initial predictions as offset. This step solves the relevant component of the efficient influence curve estimating equation.(1,6) Finally, we use the estimated coefficients to obtain targeted predictions of the expected outcome under the intervention and under the control.

# Estimate the clever covariate under the intervention A=1
# Here and below, we use H(A,W) notation in line with the larger Targeted Learning literature
h_1w <- A/pscore
# Estimate the clever covariate under the control A=0
h_0w <- (1-A)/(1-pscore)
# fluctuation regression where `qlogis` is the logit(x)=log[x/(1-x)] function in *R*
updated_fit <-  glm(Y ~ -1 +offset(qlogis(initial_prediction_A)) + h_0w + h_1w, family="binomial")
 # extract the estimated coefficients for the clever covariates

```r
 # and denote them `eps_0w` and `eps_1w`, respectively
eps_0w <- updated_fit$coef['h_0w']
eps_1w <- updated_fit$coef['h_1w']

# obtain targeted prediction  under the intervention and under the control
# here `plogis` denotes the inverse-logit function
target_prediction_1 <- plogis( qlogis(initial_prediction_1) + eps_1w / pscore )
target_prediction_0 <- plogis( qlogis(initial_prediction_0) + eps_0w / (1-pscore) )
```

4. <u>Obtain a point estimate by averaging the targeted predictions in each arm and contrasting on the scale of interest</u>. Averaging the targeted predictions corresponds to standardizing them over the covariate distribution. Therefore, we obtain a marginal effect estimate, even though we have adjusted for covariates. The interpretation of the results is equivalent to the unadjusted estimator.

```r
# Point estimates
# Psi(1) = E[E(Y|A=1,W)]
psi_1 <- mean(target_prediction_1)
# Psi(0) = E[E(Y|A=0,W)]
psi_0 <- mean(target_prediction_0)
# Absolute effect
psi_absolute <- psi_1 - psi_0
# Relative effect
psi_relative <- psi_1 / psi_0
```

5. <u>Obtain a variance estimate with the sample variance of the estimated influence curve (IC) scaled by sample size $N$.</u> We first estimate the IC for the arm-specific parameters $\Psi(a) = \mathbb{E}[\mathbb{E}(Y|A=a,W)]$ and then use the Delta method to obtain the IC for the effect of interest. For the relative effect, we do this on the log scale. With estimates of the ICs, we can obtain variance estimates and Wald-type confidence intervals.

```r
# Estimated ICs for the arm-specific parameters
ic_1 <- h_1w*(Y - target_prediction_1) + target_prediction_1 - psi_1
ic_0 <- h_0w*(Y - target_prediction_0) + target_prediction_0 - psi_0

# Delta Method for the absolute effect: psi_1 - psi_0
ic_absolute <- ic_1 - ic_0
# Delta method for relative effect: psi_1/psi_0
# on the log-scale: log(psi_1/psi_0) = log(psi_1) - log(psi_0)
ic_relative <- 1/psi_1*ic_1 - 1/psi_0*ic_0
```

```r
    # quick function to create 95% ci
    get_inference <- function(psi, ic, N, relative=F){
      # standard error estimate: square root of sample variance of the ICs, scaled by sample size
      se <- sqrt(var(ic)/N)
      # create 95% CI
      cutoff <- qnorm(0.05/2, lower.tail=F)
      # 95% confidence interval
      ci_lo <- (psi - cutoff*se)
      ci_hi <- (psi + cutoff*se)
      if(relative){
        # transform back
        psi<- exp(psi)
        ci_lo <- exp(ci_lo)
        ci_hi <- exp(ci_hi)
      }
      data.frame(psi_hat=psi, ci_lo, ci_hi)
    }

    # point estimate & 95%CI for treatment-specific parameters
    get_inference(psi_1, ic_1, N)
    get_inference(psi_0, ic_0, N)
    # for the absolute effect
    get_inference(psi_1 - psi_0, ic_absolute, N)
    # for the relative effect - on the log scale
    get_inference(log(psi_1/psi_0), ic_relative, N, relative=T)
```

We refer the reader to Moore and van der Laan for a detailed discussion of TMLE in randomized trials, its relation to G-computation and other doubly-robust estimators, alternative targeting approaches, as well as the benefits of estimating the known propensity score.(1) In particular, if the outcome regression contains an intercept and a main term for the trial arm, then the targeting step will not yield an update *when* using the known randomization probability or an unadjusted estimator (i.e. the empirical mean) for the propensity score. When the outcome regression is misspecified (as is common), incorporating covariate adjustment when estimating the propensity score improves efficiency.(4) This is shown practically for several of the trials showcased in Table 1 of the main text. As shown in Figure 2 in the main text, our procedure incorporates "collaborative" estimation of the propensity score; we select the best propensity score estimator (with or without covariate adjustment) in response to the previously selected estimator for the outcome regression. In other words, we only incorporate covariate adjustment when estimating the propensity score if it further improves empirical efficiency.

**eAppendix B: Accounting for clustering**

In cluster-randomized controlled trials (CRTs), such as SEARCH, we must account for the dependence of participants throughout: when specifying the causal model and causal estimand, when assessing identifiability, when obtaining a point and variance estimate for the corresponding statistical estimand, and when interpreting the results (e.g., (7–10)). In particular, the cross-validation scheme must respect the independent unit. Therefore, all participants from a given cluster must be assigned to the same sample-split (i.e., fold). Additionally, a cluster-level influence curve – often obtained by aggregating individual-level counterparts within clusters — is used for variance estimation.(7,10,11)

For further discussion of causal estimands and estimation approaches in CRTs, we refer the reader to Benitez et al.(10) For further discussion of Two-Stage TMLE to minimize bias and maximize power in the analysis of CRTs, we refer the reader to Balzer et al.(7) Application of Two-Stage TMLE has meaningfully changed inferences in several analyses, including SEARCH-TB where it reversed trial conclusions.(12,7,13–18)

**eAppendix C: Additional Details on the Real-Data Analyses**

For the trials in Table 1 of the main text, we provide the key ingredients for implementing Adaptive Pre-specification: the estimand (which implies the loss function as the estimated influence curve-squared for the corresponding TMLE), the candidate covariates and candidate algorithms for estimating the outcome regression and the propensity score, and the cross-validation scheme. The full Statistical Analysis Plans are linked in the corresponding paper. We also provide the point estimates from TMLE with Adaptive Pre-specification (primary analysis) and the unadjusted estimator.

- Puryear, et al. *JIAS*, 2023
    - Estimand: Relative risk (RR)
    - Covariates: sex, age, country, baseline viral suppression, baseline level of alcohol use via biomarker, baseline alcohol use via self-report, or nothing (unadjusted)

- Algorithms: working GLMs adjusting for up to 1 covariate, stepwise regression, multivariate adaptive regression splines (MARs), MARs after correlation based screening, or unadjusted
- Cross-validation: 10-fold & standard variance estimation
- Point estimates:
  - Viral suppression: RR=1.01 with TMLE & RR=1.01 with the unadjusted
  - Alcohol use: RR=0.86 with TMLE & RR=0.84 with the unadjusted
- Ayieko, et al. *JAIDS*, 2024
  - Estimand: Relative risk
  - Covariates: sex, age, country, enrollment group, mobility, baseline viral suppression, or nothing (unadjusted)
  - Algorithms: working GLMs adjusting for up to 1 covariate, stepwise regression, MARs, MARs after correlation based screening, or unadjusted
  - Cross-validation: 10-fold & standard variance estimation
  - Point estimates:
    - Viral suppression: RR=0.99 with TMLE & RR=0.97 with the unadjusted
    - Treatment possession: RR=1.07 with TMLE & RR=1.07 with the unadjusted
- Hickey, et al. *PLoS One*, 2022.
  - Estimand: Relative risk
  - Covariates: sex, age group, baseline hypertension severity, site, or nothing (unadjusted)
  - Algorithms: working GLMs adjusting for up to 1 covariate, working GLM adjusting for all covariates as main terms, stepwise regression, stepwise regression with pairwise interactions, penalized regression, or unadjusted
  - Cross-validation: 10-fold & standard variance estimation
  - Point estimate:
    - Linkage: RR=1.45 with TMLE & RR=1.46 with the unadjusted
    - Hypertension control: RR=1.23 with TMLE & RR=1.23 with the unadjusted
- Hickey, et al. *PLoS Med*, 2025
  - Estimand: Average treatment effect (ATE)
  - Covariates: sex, age, country, baseline hypertension severity, or nothing (unadjusted)

- ○ Algorithms: working GLMs adjusting for up to 1 covariate, working GLM adjusting for all covariates as main terms, stepwise regression, or unadjusted
  - ○ Cross-validation: 10-fold & standard variance estimation
  - ○ Point estimates:
    - ■ Hypertension control (in terms of risk difference [RD]): RD=26% with TMLE & RD=25% with the unadjusted
    - ■ Systolic blood pressure (in terms of difference in means): ATE=-8.2 mmHG with TMLE & ATE=-8.5 mmHG with the unadjusted estimator
- Kakande, et al. *JIAS*, 2023
  - ○ Estimand: Average treatment effect
  - ○ Covariates: sex, age, country, prior prevention use, or nothing (unadjusted)
  - ○ Algorithms: working GLMs adjusting for up to 1 covariate, stepwise regression, MARs, or unadjusted
  - ○ Cross-validation: leave-one-one-out, accounting for clustering, & standard variance estimation. There were 16 clusters.
  - ○ Point estimates:
    - ■ Prevention use: ATE=27% with TMLE & ATE=28% with the unadjusted
    - ■ Use during risk periods: ATE=36% with TMLE & ATE=36% with the unadjusted
- Kabami, et al. *JAIDS*, 2024
  - ○ Estimand: Average treatment effect
  - ○ Covariates: pregnancy status, age, country, prior prevention use, or nothing (unadjusted)
  - ○ Algorithms: working GLMs adjusting for up to 1 covariate, stepwise regression, MARs, or unadjusted
  - ○ Cross-validation: 10-fold & standard variance estimation
  - ○ Point estimates:
    - ■ Prevention use: ATE=40% with TMLE & ATE=40% with the unadjusted
    - ■ Use during risk periods: ATE=38% with TMLE & ATE=37% with the unadjusted
- Koss, et al. *AIDS*, 2024
  - ○ Estimand: Average treatment effect
  - ○ Covariates: sex, age, country, prior prevention use, or nothing (unadjusted)

- Algorithms: working GLMs adjusting for up to 1 covariate, stepwise regression, MARs, or unadjusted
    - Cross-validation: 10-fold & standard variance estimation
    - Point estimates:
        - Prevention use: ATE=29% with TMLE & ATE=31% with the unadjusted
        - Use during risk periods: ATE=39% with TMLE & ATE=40% with the unadjusted
- Kamya, et al. *Lancet HIV*, 2024: This was a randomized trial extension, which combined three prior trials. Therefore, the primary analysis adjusted indicator for trial in the outcome regression. Our machine learning procedure was applied to select additional (if any) adjustment variables.
    - Estimand: Average treatment effect
    - Covariates: sex, age, baseline alcohol use, baseline mobility, or no further adjustment
    - Algorithms: working GLMs adjusting for up to 1 additional covariate, penalized regression (LASSO), or no further adjustment
    - Cross-validation: leave-one-one-out, accounting for clustering, & standard variance estimation. There were 625 clusters.
    - Point estimates:
        - Prevention use: ATE=56% with TMLE & ATE=57% with the unadjusted
        - Use during risk periods: ATE=60% with TMLE & ATE=61% with the unadjusted

**eAppendix D: Implications for Power Calculations**

Throughout, we have emphasized how our adaptive procedure is guaranteed not to harm precision and has meaningfully improved precision in several trials. As such, we are often asked if sample size and power calculations should be based on the unadjusted effect estimator or an approach incorporating covariate adjustment. Our recommendation is to base power calculations on the standard formulas using unadjusted effect estimators. These calculations will be conservative if an approach using covariate adjustment is selected for the analysis. However, these calculations will *not* be anti-conservative if an unadjusted approach is selected for the trial analysis or if the assumed precision gains from covariate adjustment are too optimistic. In other words, we would rather be over-powered than under-powered. We note, however, that others have advocated for conducting power calculations based on approaches using covariate adjustment.(19)

We again reiterate our recommendation to conduct simulations mimicking your trial. Such simulations can highlight the potential gains in precision from covariate adjustment. Furthermore, they can help inform specification of the Statistical Analysis Plan, confirm implementation of the computing code, and verify Type-I error control (under the null).